\documentclass[prl,reprint]{revtex4-1}
\pdfoutput=1
\usepackage{graphicx}
\usepackage{bm}
\usepackage{hyperref}

\begin{document}

\title{Evidence for a $\nu=5/2$ Fractional Quantum Hall Nematic State in Parallel Magnetic Fields}
\date{today}

\author{Yang Liu}
\affiliation{Department of Electrical Engineering,
Princeton University, Princeton, New Jersey 08544}
\author{S.\ Hasdemir}
\affiliation{Department of Electrical Engineering,
Princeton University, Princeton, New Jersey 08544}
\author{M.\ Shayegan}
\affiliation{Department of Electrical Engineering,
Princeton University, Princeton, New Jersey 08544}
\author{L.N.\ Pfeiffer}
\affiliation{Department of Electrical Engineering,
Princeton University, Princeton, New Jersey 08544}
\author{K.W.\ West}
\affiliation{Department of Electrical Engineering,
Princeton University, Princeton, New Jersey 08544}
\author{K.W.\ Baldwin}
\affiliation{Department of Electrical Engineering,
Princeton University, Princeton, New Jersey 08544}

\date{\today}

\begin{abstract}

  We report magneto-transport measurements for the fractional quantum
  Hall state at filling factor $\nu=$ 5/2 as a function of applied
  parallel magnetic field ($B_{||}$). As $B_{||}$ is increased, the
  5/2 state becomes increasingly anisotropic, with the in-plane
  resistance along the direction of $B_{||}$ becoming more than 30
  times larger than in the perpendicular direction. Remarkably, the
  resistance anisotropy ratio remains constant over a relatively large
  temperature range, yielding an energy gap which is the same for both
  directions. Our data are qualitatively consistent with a fractional
  quantum Hall \textit{nematic} phase.

\end{abstract}


\maketitle

The origin and properties of the fractional quantum Hall state (FQHS)
at the even-denominator Landau level (LL) filling factor $\nu = $ 5/2
\cite{Willett.PRL.1987} have become of tremendous current
interest. This is partly because the quasi-particle excitations of the
5/2 FQHS are expected to obey non-Abelian statistics
\cite{Moore.Nuc.Phy.1991} and be useful for topological quantum
computing \cite{Nayak.Rev.Mod.Phys.2008}. The stability and robustness
of the 5/2 state, and its sensitivity to the parameters of the hosting
two-dimensional electron system (2DES) are therefore of paramount
importance. This stability has been studied as a function of 2DES
density, quantum well width, disorder, and a parallel magnetic field
($B_{||}$) applied in the 2DES plane \cite{Eisenstein.PRL.1988,
  Pan.PRL.1999.Stripe, Lilly.PRL.83.1999, Pan.PRB.2008, Dean.PRL.2008,
  *Dean.PRL.2008.101, Choi.PRB.2008, Nuebler.PRB.2010,
  Shabani.PRL.2010, Xia.PRL.2010, Kumar.PRL.2010, Zhang.PRL.2010,
  Pan.PRL.2011, Liu.PRL.2011, Gamez.cond.mat.2011,
  Samkharadze.PRB.2011}. Recent measurements made as a function of
density and quantum well width, for example, have demonstrated that
the $\nu=5/2$ FQHS is stable when the Fermi level lies in the
excited-state ($N=1$) LL of the symmetric electric subband but turns
into a compressible state if the Fermi level moves to the ground-state
($N=0$) LL of the anti-symmetric electric subband
\cite{Shabani.PRL.2010,Liu.PRL.2011}. The role of $B_{||}$ is also
important and has been used to shed light on the spin polarization of
the 5/2 state, which in turn has implications for whether or not the
state is non-Abelian \cite{Eisenstein.PRL.1988, Dean.PRL.2008,
  Zhang.PRL.2010}. The application of $B_{||}$ in fact has more subtle
consequences. It often induces anisotropy in the 2DES transport
properties in the $N=1$ LL and, at sufficiently large values of
$B_{||}$, leads to an eventual destruction of the $\nu=5/2$
FQHS \cite{Pan.PRL.1999.Stripe, Lilly.PRL.83.1999, Xia.PRL.2010},
replacing it by a compressible, anisotropic ground state. This
anisotropic state is reminiscent of the non-uniform density, stripe
phases seen at half-integer fillings in the higher ($N>1$)
LLs \cite{Lilly.PRL.1999, Du.SSC.1999}.


Here we study the $\nu=5/2$ FQHS as a function of $B_{||}$ in a very
high-quality 2DES. We find that the application of $B_{||}$ leads to a
strong anisotropy in transport as the resistance along $B_{||}$
becomes more than 30 times larger than in the perpendicular
direction. Nevertheless, at low temperatures ($T\lesssim 100$ mK),
the resistances along these two in-plane directions monotonically
decrease with decreasing temperature while the anisotropy ratio
remains nearly constant. From the temperature-dependence of the
resistances, we are able to measure the energy gap ($\Delta$) for the
5/2 FQHS along the two in-plane directions. Despite the enormous
transport anisotropy, $\Delta$ has the same magnitude along both
directions. We interpret our data in terms of a FQH \textit{nematic}
phase.

In our sample, which was grown by molecular beam epitaxy, the 2DES is
confined to a 30-nm-wide GaAs quantum well, flanked by undoped
Al$_{0.24}$Ga$_{0.76}$As spacer layers and Si $\delta$-doped
layers. The 2DES has a density of $n=$ 3.0 $\times 10^{15}$ m$^{-2}$
and a very high mobility, $\mu \simeq$ 2,500 m$^2$/Vs. It has a very
strong $\nu=5/2$ FQHS, with an energy gap of $\Delta\simeq 0.4$ K,
when $B_{||}=0$. The sample is 4 mm $\times$ 4 mm with alloyed InSn
contacts at four corners. For the low-temperature measurements, we
used a dilution refrigerator with a base temperature of $T \simeq$ 20
mK, and a sample platform which could be rotated \textit{in-situ} in
the magnetic field to induce a parallel field component $B_{||}$ along
the $x$-direction. We use $\theta$ to express the angle between the
field and the normal to the sample plane, and denote the longitudinal
resistances measured along and perpendicular to the direction of
$B_{||}$ as $R_{xx}$ and $R_{yy}$, respectively.

\begin{figure*}
\includegraphics[width=.95\textwidth]{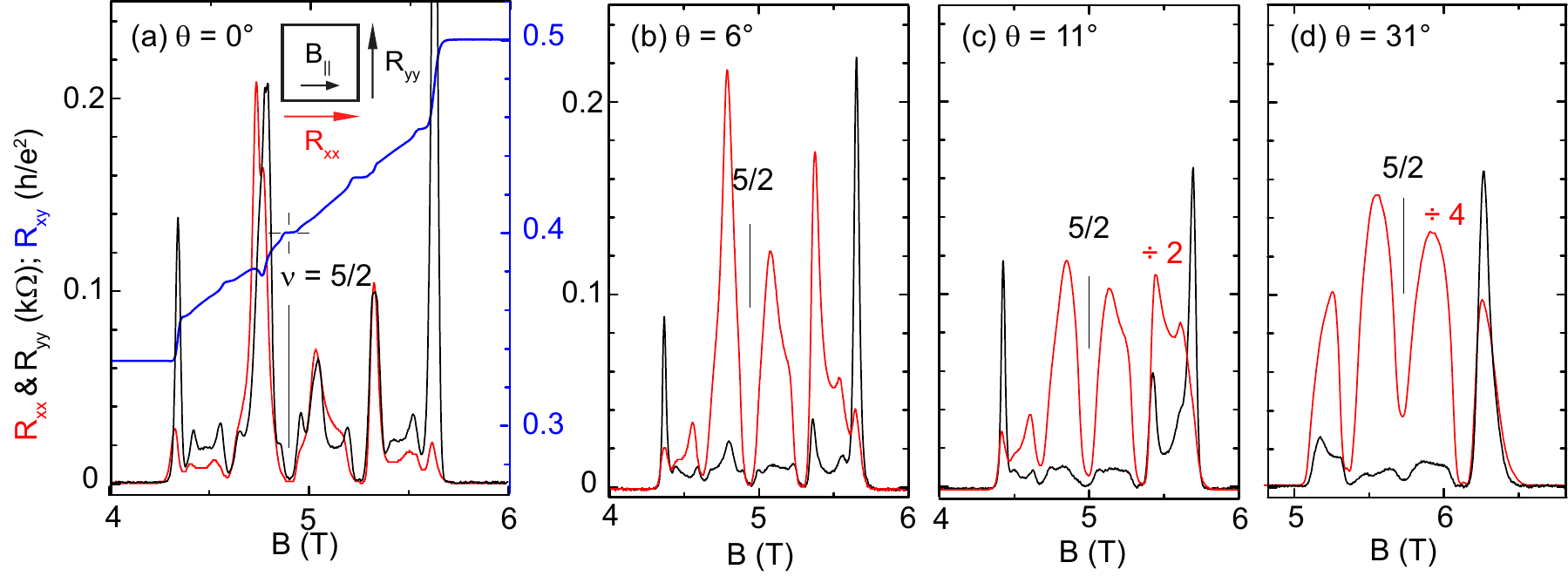}
\caption{\label{fig:cartoon}(color online) (a) Longitudinal
  resistances $R_{xx}$ (red) and $R_{yy}$ (black), and Hall resistance
  $R_{xy}$ (blue) measured as a function of perpendicular magnetic
  field. The deep minima in $R_{xx}$ and $R_{yy}$, as well as the
  well-quantized $R_{xy}$ plateau, indicate a strong FQHS at
  $\nu=5/2$. (b-d) $R_{xx}$ and $R_{yy}$ measured at finite tilting
  angles, $\theta=6^{\circ}$, $11^{\circ}$ and $31^{\circ}$ are shown
  as a function of total magnetic field. The in-plane component of the
  magnetic field ($B_{||}$) is along the $x$-direction. Note that the
  $R_{xx}$ traces in (c) and (d) are divided by factors of 2 and
  4. Strong transport anisotropy near $\nu=5/2$ grows as $\theta$
  increases. All traces were recorded at the base temperature of our
  measurements, $T \simeq$ 20 mK.}
\end{figure*}

\begin{figure*}
\includegraphics[width=.95\textwidth]{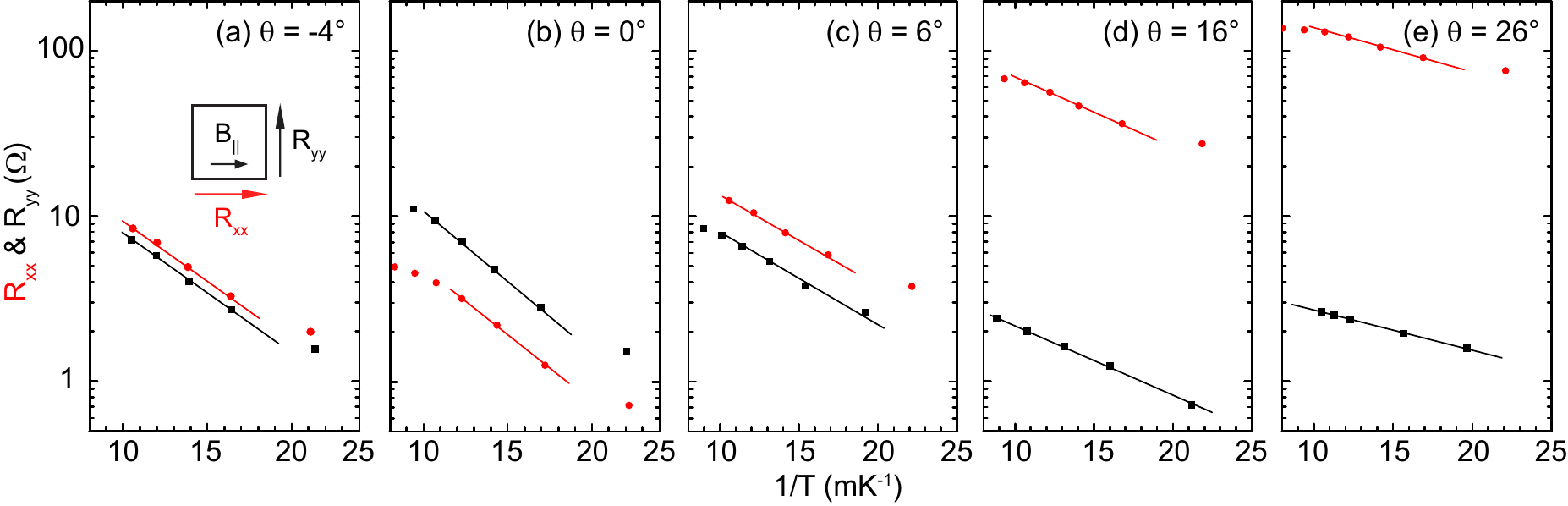}
\caption{\label{fig:waterfall} (color online) Temperature-dependence
  of $R_{xx}$ (red) and $R_{yy}$ (black) at $\nu=5/2$, measured at
  different tilting angles, $\theta$. The excitation gap deduced from
  the slopes of these plots decreases as $\theta$ is increased, while
  the transport anisotropy increases.}
\end{figure*}

\begin{figure}[htb]
\includegraphics[width=.48\textwidth]{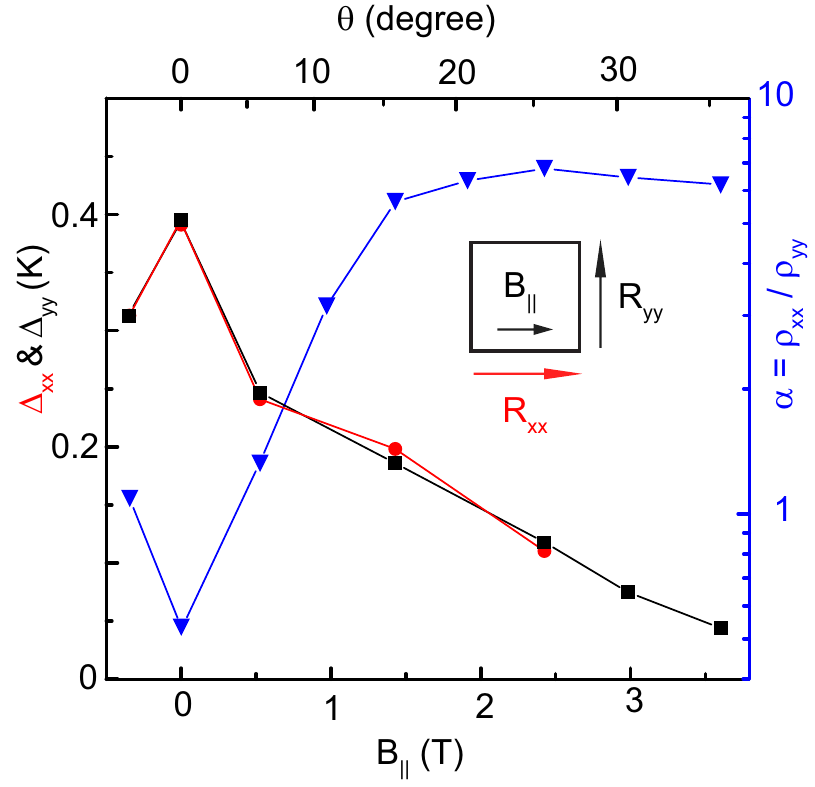}%
\caption{\label{fig:colorful} (color online) Measured excitation gaps,
  $\Delta_{xx}$ (red circles) and $\Delta_{yy}$ (black squares), are
  shown as a function of the in-plane magnetic field
  $B_{||}$. $\Delta_{xx}$ and $\Delta_{yy}$ nearly equal each other
  and decrease with increasing $B_{||}$. Also plotted (blue triangles
  is the transport anisotropy factor, $\alpha$, defined as the ratio
  between the resistivities $\rho_{xx}$ and $\rho_{yy}$). Note the
  logarithmic scale on the right: $\alpha$ grows exponentially with
  $B_{||}$ at small $B_{||}$, and saturates at large $B_{||}\gtrsim 1$
  T.}
\end{figure}

\begin{figure}
\includegraphics[width=.48\textwidth]{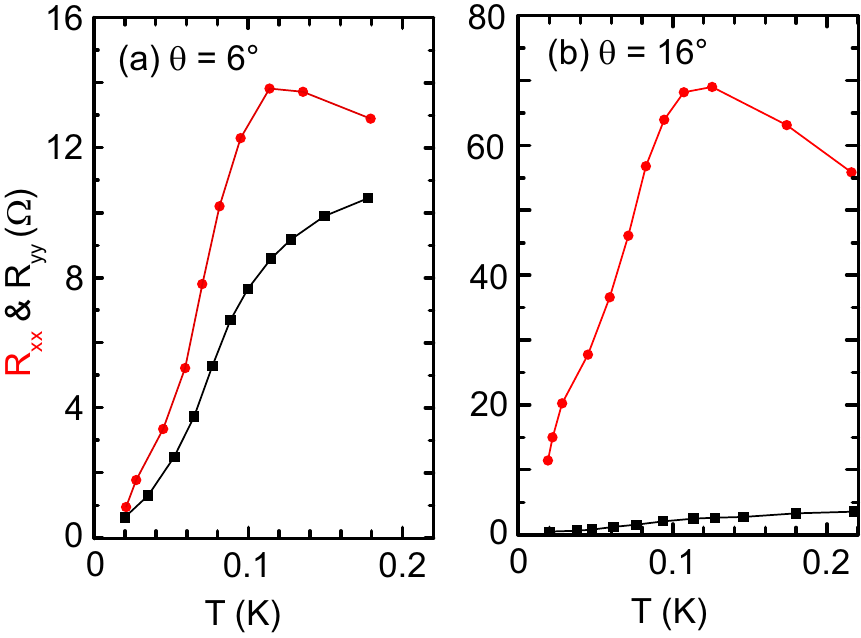}
\caption{\label{fig:waterfall2}(color online) $R_{xx}$ and $R_{yy}$
  are shown vs. temperature for two values of tilt angle. $R_{yy}$
  monotonically increases with increasing temperature, while $R_{xx}$
  shows a downturn at high temperatures $T\gtrsim 120$ mK, indicating
  a smaller anisotropy.}
\end{figure}

Figure 1 shows the $R_{xx}$ (red) and $R_{yy}$ (black) measured as a
function of the total magnetic field in the filling range $2<\nu<3$;
the Hall resistance $R_{xy}$ is also shown (in blue) in Fig. 1(a). The
traces in Fig. 1(a) were taken at $\theta=0$, i.e., for $B_{||}=0$,
and exhibit a very strong $\nu=5/2$ FQHS with an energy gap of
$\simeq$0.4 K and an $R_{xy}$ which is well-quantized at
0.4$h/e^2$. As seen in Figs. 1(b-d), the application of $B_{||}$
causes a very pronounced anisotropy in the in-plane transport at and
near $\nu=5/2$, and $R_{xx}$ becomes much larger than $R_{yy}$. At
$\theta=26^{\circ}$, e.g., $R_{xx}$ is about 30 times $R_{yy}$. Note
that in our experiments $B_{||}$ is applied along the $x$-direction so
that the "hard"-axis we observe for in-plane transport is along the
direction of $B_{||}$. This is consistent with previous reports on
$B_{||}$-induced resistance anisotropy near 5/2
\cite{Pan.PRL.1999.Stripe,Lilly.PRL.83.1999, Xia.PRL.2010, Note1}.


In Fig. 2 we show the temperature dependence of $R_{xx}$ and $R_{yy}$
at $\nu=5/2$ for different values of $\theta$. In the temperature
range $50<T<100$ mK, both $R_{xx}$ and $R_{yy}$ are activated and
follow the relation $R\sim \exp(-\Delta/2k_BT)$, where $\Delta$ is FQH
energy gap. At $\theta=0$ $R_{yy}$ is larger than $R_{xx}$ by about a
factor of two. This anisotropy is caused by a mobility anisotropy, as
the latter is often seen in very high-mobility samples. With the
application of a very small $B_{||}$ along the $x$-direction
($|\theta|\lesssim 5^{\circ}$), the anisotropy reverses so that
$R_{xx}$ exceeds $R_{yy}$. This trend continues with increasing
$\theta$ and, at $\theta=26^{\circ}$, $R_{xx}$ becomes 30 times larger
than $R_{yy}$. Remarkably, however, despite the very large anisotropy,
both $R_{xx}$ and $R_{yy}$ remain activated and yield very similar
values for $\Delta$.

The transport energy gaps at $\nu=5/2$ measured as a function of
$B_{||}$ up to $\simeq 3.6$ T are summarized in Fig. 3. We denote the
energy gaps deduced from the temperature-dependence of $R_{xx}$ and $R_{yy}$
by $\Delta_{xx}$ and $\Delta_{yy}$, respectively. It is clear in
Fig. 3 that $\Delta_{xx}\simeq\Delta_{yy}$ despite the large
anisotropy observed in $R_{xx}$ and $R_{yy}$. In Fig. 3 we also plot
the observed transport anisotropy as a function of $B_{||}$. Here we
used the values of $R_{xx}$ and $R_{yy}$ resistances at $T=60$ mK,
converted them to \textit{resistivities} $\rho_{xx}$ and $\rho_{yy}$
following the formalism presented in Ref. \cite{Simon.PRL.1999}, and
plot the ratio $\alpha=\rho_{xx}/\rho_{yy}$. As a function of
$B_{||}$, this ratio grows very quickly, approximately exponentially
up to $B_{||} \simeq 1$ T, and then saturates at higher $B_{||}$. The
energy gaps $\Delta_{xx}$ and $\Delta_{yy}$, however, exhibit a very
steep drop at small $B_{||}$, followed by a more gradual and monotonic
decrease at higher $B_{||}$. For $\theta>36^{\circ}$ ($B_{||} \gtrsim
3.6$ T) we cannot measure the gap for the 5/2 FQHS as it becomes too
weak.

The data presented above provide clear evidence for a strong $\nu=5/2$
FQHS whose in-plane transport is very anisotropic in the presence of
applied $B_{||}$. And, remarkably, its energy gap is the same for the
two in-plane directions. These observations imply a $\nu=5/2$ FQHS
whose transport is anisotropic at finite temperatures. A possible
interpretation of our data is that we are observing a FQH
\textit{nematic} phase. It has been argued in numerous theoretical
studies that such liquid-crystal-like FQH phases might exist in 2D
systems where the rotational symmetry is broken
\cite{Balents.EPL.1996, Musaelian.JPCM.1996, Fradkin.PRB.1999,
  Fogler.EPL.2004,Mulligan.PRB.2011, Haldane.PRL.2011, Qiu.PRB.2012,
  BoYang.PRB.2012,HaoWang.PRB.2012}. We note that in a 2DES with
finite (non-zero) electron layer thickness, such as ours, $B_{||}$
breaks the rotational symmetry as it couples to the electrons'
out-of-plane motion and causes an anisotropy of their real-space
motion as well as their Fermi contours
\cite{Kamburov.PRB.2012}. Recently it was indeed demonstrated
experimentally that such an anisotropy is qualitatively transmitted to
the quasiparticles at high magnetic fields, for example to composite
Fermions \cite{Kamburov.preprint.2013}. It is therefore plausible that
$B_{||}$ which breaks the rotational symmetry in our 2DES would lead a
FQH nematic phase at $\nu=5/2$.

A FQH nematic phase was in fact recently proposed theoretically
\cite{Mulligan.PRB.2011} to explain the experimental observations of
Xia \textit{et al.} \cite{Xia.Nat.Phys.2011} for another FQHS in the
$N=1$ LL, namely at $\nu=7/3$. In the model of
Ref. \cite{Mulligan.PRB.2011}, the ground-state is a FQHS but the dc
longitudinal resistance at finite temperatures is anisotropic as it
reflects the anisotropic property of the thermally excited
quasiparticles. The energy gap for the excitations, however, is
predicted to be the same for $R_{xx}$ and $R_{yy}$. These features are
consistent with our experimental data. According to Mulligan
\textit{et al.}, the FQH nematic phase with anisotropic transport is
stable only at very low temperatures\cite{Mulligan.PRB.2011}. As
temperature is raised above a critical value that depends on the
details of the sample's parameters and transport properties, $R_{xx}$
should abruptly drop and $R_{yy}$ suddenly rise so that they have the
same value, signaling an isotropic FQH phase. Mulligan \textit{et al.}
also report that, thanks to the small symmetry-breaking $B_{||}$
field, this finite-temperature transition might become rounded so that
$R_{xx}$ and $R_{yy}$ approach each other more slowly at high
temperatures (see Fig. 3 of Ref. \cite{Mulligan.PRB.2011}). As
mentioned above, our data at low temperatures are qualitatively
consistent with the predictions of Ref. \cite{Mulligan.PRB.2011} for
the FQH nematic state. At higher temperatures (Fig. 4), our data
exhibit a downturn in $R_{xx}$ as temperature is raised above $\simeq$
0.1 K, signaling that transport is becoming less anisotropic, also
generally consistent with Ref. \cite{Mulligan.PRB.2011}
predictions. However, up to the highest temperatures achieved in our
measurements ($\simeq$ 0.2 K, which is comparable to the excitation
gap), we do not see a transition to a truly isotropic state.


While the above interpretation of our data based on a FQH nematic
state is plausible, there might be alternative explanations. For
example, it has been theoretically suggested that the low-energy
charged excitations of the FQHSs in the $N=1$ LL have a very large
size as they are complex composite Fermions dressed by roton clouds
\cite{Balram.preprint.2013}. Because of their large size, these
excitations are prone to become anisotropic in the presence of
$B_{||}$. Such anisotropy, even if small in magnitude, could lead to a
much larger \textit{transport} anisotropy of the quasiparticle
excitations at finite temperatures because this transport would
involve hopping or tunneling of the quasiparticles between the
localized regions. 


To summarize, our magneto-transport measurements reveal that the
application of a $B_{||}$ leads to a $\nu=5/2$ FQHS whose in-plane
longitudinal resistance is highly anisotropic at low temperatures. The
resistance anisotropy ratio remains constant over a relatively large
temperature range, and the energy gap we extract from the
temperature-dependence of the resistances is the same for both
directions. Our data are generally consistent with a fractional
quantum Hall \textit{nematic} phase, although other explanations might
be possible. Regardless of the interpretations, our results attest to
the very rich and yet not fully understood nature of the enigmatic
$\nu=5/2$ FQHS.

\begin{acknowledgments}
  We acknowledge support from the the NSF (DMR-0904117 and MRSEC
  DMR-0819860), and Moore and Keck Foundations. A portion of this work
  was performed at the National High Magnetic Field Laboratory, which
  is supported by NSF Cooperative Agreement No. DMR-0654118, by the
  State of Florida, and by the DOE. We thank R. Bhatt, J. K. Jain,
  D. Haldane, and Z. Papic for illuminating discussions, and E. Palm,
  J. H. Park, T. P. Murphy and G. E. Jones for technical assistance.
\end{acknowledgments}

\bibliography{../bib_full}
\end{document}